\documentclass[11pt,twoside]{article}
\usepackage{./asp2014}
\begin{document}
\resetcounters

\bibliographystyle{asp2014}

\markboth{Rich et al.}{A New Kinematic Survey of the Bulge AGB}


\title{From BRAVA to BAaDE: A New Kinematic Survey \\ of the Bulge AGB}
\author{R. Michael Rich,$^1$ Adam Trapp,$^1$ M. R. Morris,$^1$ L. Sjouwerman,$^2$ M. Stroh,$^3$ M. Claussen,$^2$ and Y. Pihlstrom$^{2,3}$}
\affil{$^1$Department of Physics and Astronomy, UCLA, Los Angeles, CA 90095, USA; \email{rmr@astro.ucla.edu, atrapp@astro.ucla.edu, morris@astro.ucla.edu}}
\affil{$^2$ National Radio Astronomy Observatory, 1003 Lopezville Rd, Socorro, NM 87801, USA; \email{lsjouwer@nrao.edu, mclausse@nrao.edu, }}
\affil{$^3$ Department of Physics and Astronomy, University of New Mexico, Albuquerque, NM 87131, USA; \email{ylva@unm.edu, mstroh@unm.edu }}

\paperauthor{R. Michael Rich}{rmr@astro.ucla.edu}{}{UCLA}{Department of Physics and Astronomy}{Los Angeles}{California}{90095}{USA}
\paperauthor{Adam Trapp}{atrapp@astro.ucla.edu}{}{UCLA}{Department of Physics and Astronomy}{Los Angeles}{California}{90095}{USA}
\paperauthor{M. R. Morris}{morris@astro.ucla.edu}{}{UCLA}{Department of Physics and Astronomy}{Los Angeles}{California}{90095}{USA}
\paperauthor{L. Sjouwerman}{lsjouwer@nrao.edu}{}{NRAO}{}{Socorro}{New Mexico}{87801}{USA}
\paperauthor{M. Claussen}{mclausse@nrao.edu}{}{NRAO}{}{Socorro}{New Mexico}{87801}{USA}
\paperauthor{M. Stroh}{mstroh@unm.edu}{}{University of New Mexico}{Department of Physics and Astronomy}{Albuquerque}{New Mexico}{87131}{USA}
\paperauthor{Y. Pihlstrom}{ylva@unm.edu}{}{University of New Mexico}{Department of Physics and Astronomy}{Albuquerque}{New Mexico}{87131}{USA}

\begin{abstract}

We consider interesting results on the kinematics of the bulge, beginning with Mould's (1983) measurement of the velocity dispersion of the late M giants in the bulge, and carrying the discussion on to the present day, to discuss the largest survey ever undertaken of luminous SiO maser AGB stars.   This SiO maser population of AGB stars is too luminous to have evolved from the oldest globular cluster age stellar population, yet has a  velocity dispersion consistent with the oldest populations (the red clump stars) in the bulge.  We identify two groups of maser stars by their kinematics; the subgroup with kinematics similar to the bulge has the highest bolometric luminosities with some giants reaching $M_{bol}<-6$.  In a new result, we find that the most luminous maser stars with bulge kinematics are concentrated to the Galactic plane with the most luminous maers being found at $b<\mid 1.5^\circ \mid$; while the disk maser population shows no such difference. This mirrors a similar concentration to the Galactic Center found for long period Miras, noted in earlier studies.  We conclude that the maser population represents another facet of the growing tension between evidence supporting a bulge population dominated by old stars, and a significant fraction of intermediate age stars.  

\end{abstract}

\section{Introduction}

Perhaps because of our imperfect perspective on the Galaxy's center as seen from the Northern hemisphere, and of course, the 30 magnitudes of optical extinction toward the center of the Milky Way, our views of the Galactic Center have historically somehow missed the mark.  One of the earliest studies of the Galactic center could properly be said to be Shapley's use of RR Lyrae in globular clusters to prove that our Sun is at the edge rather than the center of the Milky Way.  Baade's development of two stellar populations was crucial, but his discovery of RR Lyrae stars in the Galactic Center would leave some generations of astronomers improperly describing the bulge as ancient and metal-poor (like the globular clusters) because of Baade's (1951) momentous identification of RR Lyrae stars in Baade's Window, which indeed also gave a remarkably precise distance to the Center of 8 kpc.  In fact, the incorrect description of the bulge as being old and metal-poor would persist well into the 1970s, even with Arp's (1962) pioneering color-magnitude diagram of Baade's Window that clearly showed a wide range of metallicity.  It is indeed remarkable that {\it even with the large numbers of late-type M giants found toward the Galactic Center} (Gaposchkin 1955; Nassau \& Blanco 1958), the full impact of that M giant population (present in the bulge, but not in the globular clusters) was not appreciated by the vast majority of astronomers.  However, the implications were understood at the 1957 Vatican meeting on stellar populations.  The bulge was not assigned to the oldest population II, but in fact, assigned to the {\it old disk} --- a visionary classification.  Our present-day theory for the formation of the bulge is that the buckling of the disk into a bar was in fact a critical phase in the formation of the bulge/bar that we observe today.

The honoree of this meeting, Jeremy Mould, has made so many fundamental contributions that, as Wal Sargent was apt to say about Goldreich's contributions, they have become ``part of the furniture.''  Among those seminal papers is one less noticed, Mould's (1983) radial velocity distribution of 50 M giants in the $(l,b)=0.9^\circ, -3.9^\circ$ Baade's Window field, resulting in a value of 113$\pm 11$ km/sec for the bulge velocity dispersion---a value confirmed by Rich (1990) for the K giants, years later.  Mould's study was possible, however, because it used the Blanco, McCarthy, \& Blanco (1984) identifications of late M giants in Baade's Window.  These M giants were in fact discovered using visual classification of slitless grism spectra taken with photographic plates at the prime focus of the CTIO 4m telescope.  The finding charts are from IV-N (red sensitive) photographic plates and the positions were measured using a measuring engine.    The use of M giants to measure a velocity dispersion was daring, as the photon-counting Shectograph on the 2.5m duPont telescope was nominally more blue sensitive.  The TiO bandheads yield excellent cross-correlation peaks and the red spectral energy distribution puts most of the flux redward of 7000A where the reddening is lower.   Perhaps Mould's earlier work on TiO bands (Mould \& McElroy 1978) gave him the confidence that a valid cross-correlation peak would result even though the actual bandstrengths change very strongly with temperature, including the appearance of VO, and in the coolest stars, the Ca Triplet can be nearly veiled by the molecules.   The likelihood is that Mould (1983) inspired the larger Sharples, Walker, \& Cropper (1990) survey of M giants using the first multi-fiber spectrographs at the AAT.

Related to bulge research today is Mould's theoretical work on iron hydride
in stellar atmospheres (Mould \& Wyckoff 1978), studied at present in
elliptical galaxies to constrain the IMF, and Mould (1978) on the widely used Mg index; this latter paper was one of the first to take on a deeper understanding of the widely used measurement of metallicity in the integrated light of galaxies.   Mould \& McElroy (1978) also explore the behavior of the TiO molecule in globular cluster giants.

In this review I discuss the arc from Mould's initial study of the M giants in the bulge to the
first large-scale optical/IR survey of bulge kinematics, the  Bulge Radial Velocity Assay (BRAVA) that employed 9,500 M giants to map the rotation curve and velocity distribution of the bulge.  This review extends this work to the kinematics of the AGB SiO maser stars of the BAaDE survey, which were presented by coauthor Adam Trapp at this meeting.

This project proved to be a strong demonstration of the barlike kinematics of the bulge (Shen et al.\ 2010).   Just as the major summary paper of Kunder et al.\ (2012) appeared, a new large scale survey (ARGOS) began its publications (e.g. Ness et al.\ 2012). The two surveys contrast in a number of respects.
BRAVA is more compact and dense, covering the inner bulge and targeting only the most luminous and reddest M giants.  This was advantageous in making the spectra easy to observe; we could use either the Ca infrared triplet for cross-correlation, or the TiO bands; in fact in the case of BRAVA, the Ca triplet proved most stable and durable.  Due to the red cutoff of ${\sim}7400\AA$ in the du Pont Shectograph  (photon counting detector) spectra, Mould (1983) used only the TiO bands for cross-correlation.   It is important to emphasize that BRAVA used the CTIO 4m Hydra spectrograph, while Ness et al.\ had access to the far more advanced AAOMEGA platform, with 300 fibers over 3 square degrees.  Hence Ness et al.\ were able to contemplate pushing to the fainter K giants, and integrating long enough to measure features like the 5170 Mg lines as well as the Ca infrared triplet.  This offered abundance constraints that were not available to BRAVA.  After the fact, an attempt was made to glean abundance constraints from the  Ca IR triplet by  Kunder et al.\ (2012), but a majority of the M giants were too cool and hence lacked the Ca lines required for the calibration.

\subsection{ The Age of the Bulge:  An Enduring Controversy}

It could not be an easier problem in principle: use the Hubble Space Telescope to reach several magnitudes below the main-sequence turnoff and determine the age of the bulge.  In practice, there has been a decades-long controversy concerning the age and star formation history of the bulge.  In a nutshell, the bulge has two populations of variable stars that each tell a different story.  The Mira variable population includes very-long-period variables that are an indicator of intermediate age stellar populations (Catchpole et al.\ 2016) while  RR Lyrae variables (the gold standard for the {\it oldest} stellar populations) were known to be present as well. The RR Lyrae provided a distance to the Galactic Center derived from their luminosities, long before the discovery of the 2 $\mu$m nuclear cluster.   An excellent discussion of the implications of Mira variables for intermediate age stellar populations, and the history of study of the bulge Miras, is given in Catchpole et al.\ (2016) and see also Whitelock's review in this volume.  Also Rich (2013) gives an historical overview of the bulge; Nataf (2016) reviews the age and he considers an array of evidence in support of an intermediate-age population, from the properties of AGB stars to the planetary nebulae.

There have been numerous HST studies of the bulge main sequence turnoff.  Two broad approaches have been employed: the proper motion cleaning method (Kuijken \& Rich 2002; Clarkson et al.\ 2008) and the statistical subtraction from the bulge color-magnitude diagram of foreground disk stars (Zoccali et al.\ 2003).  Both approaches call for a mostly $>10$ Gyr old stellar population.  Complications include the severe crowding, differential reddening, broad abundance distribution, extended depth distribution of the bulge, and imperfect kinematic separation of populations based on proper motion.	However, overall, every careful study finds a population dominated by 10+ Gyr old metal-rich stars; a recent luminosity function study (analagous to that of Ortolani et al.\ 1995) finds both metal-rich and metal-poor sub-populations are ${\sim}10$ Gyr old; a recent analysis confirms this finding, placing very stringent constraints on any significant intermediate age metal-rich stellar population (Renzini et al.\ 2018).

However, there are two powerful counter-arguments to the bulge having no population younger than ${\sim}10$ Gyr.  In a series of papers culminating in Bensby et al.\ (2017), microlens-amplified, high $S/N$ spectra of dwarfs are analyzed self-consistently.  They find that the metal-rich subcomponent of the population appears to lie on intermediate age isochrones.   The method gives high precision abundance analysis of the individual spectra to overcome uncertainty in reddening and distance.   When $\log g$, $\log T_{\rm eff}$, and [Fe/H] are derived from a high $S/N$ spectrum, one in principle has extracted all of the possible physical parameters, but to what population do these lensed dwarfs really belong?  It would be better to know for each case the actual pre-amplified magnitudes.  The sample size of 100 is modest, but large enough that the findings cannot be dismissed. 

The Mira variables and luminous AGB stars pose another concern; periods longer than 500 days indicate intermediate age progenitors.  Catchpole et al.\ (2016) finds such stars across the bulge/bar; the long-period population is found mostly in the bar, while the shorter period (older) AGB stars appear to occupy a more classical spheroid. There is presently a debate about whether or not the bulge RR Lyrae follow the bar, although Kunder et al.\ (2016) clearly shows they have strikingly lower rotation and higher velocity dispersion.   The BAaDE project reviewed here also finds luminous evolved AGB stars but with a surprisingly high radial velocity dispersion.  The problem of the age distribution of the bulge has been with us for 30 years, and continues to this day.    In the review of Nataf (2016), he discusses the detection of the unstable isotope Tc in AGB stars (Uttenthaler et al.\ 2007) and various studies of PNe; because trace populations can evolve from binary star evolution (see e.g. Greggio \& Renzini 1990) it is difficult to assess the actual impact of these populations on the intermediate age star fraction.

\section{Extending Mould (1983):  The BRAVA survey}

The success of 2MASS made it possible to select late M giants across the bulge; that development combined with the Hydra multi-object spectrograph at the Blanco 4m telescope at Cerro Tololo made it feasible to construct the Bulge Radial Velocity Assay, or BRAVA.  Figure~1 shows our full reddening-corrected RGB sample and the selected fields.  This study gave the first densely sampled stellar kinematic survey of the high surface brightness area of the {\it COBE} bar, and the data made possible the Portail et al.\ (2015) estimate of the bulge mass based on the kinematics of the stars.   Figure \ref{bravaresults} shows the velocity field of BRAVA.  BRAVA also was the first study to find ``cylindrical'' or pattern rotation (Howard et al.\ 2008) and the kinematics are most consistent with a bar-dominated bulge (Shen et al.\ 2010).  

\articlefigure[width=1.1\textwidth]{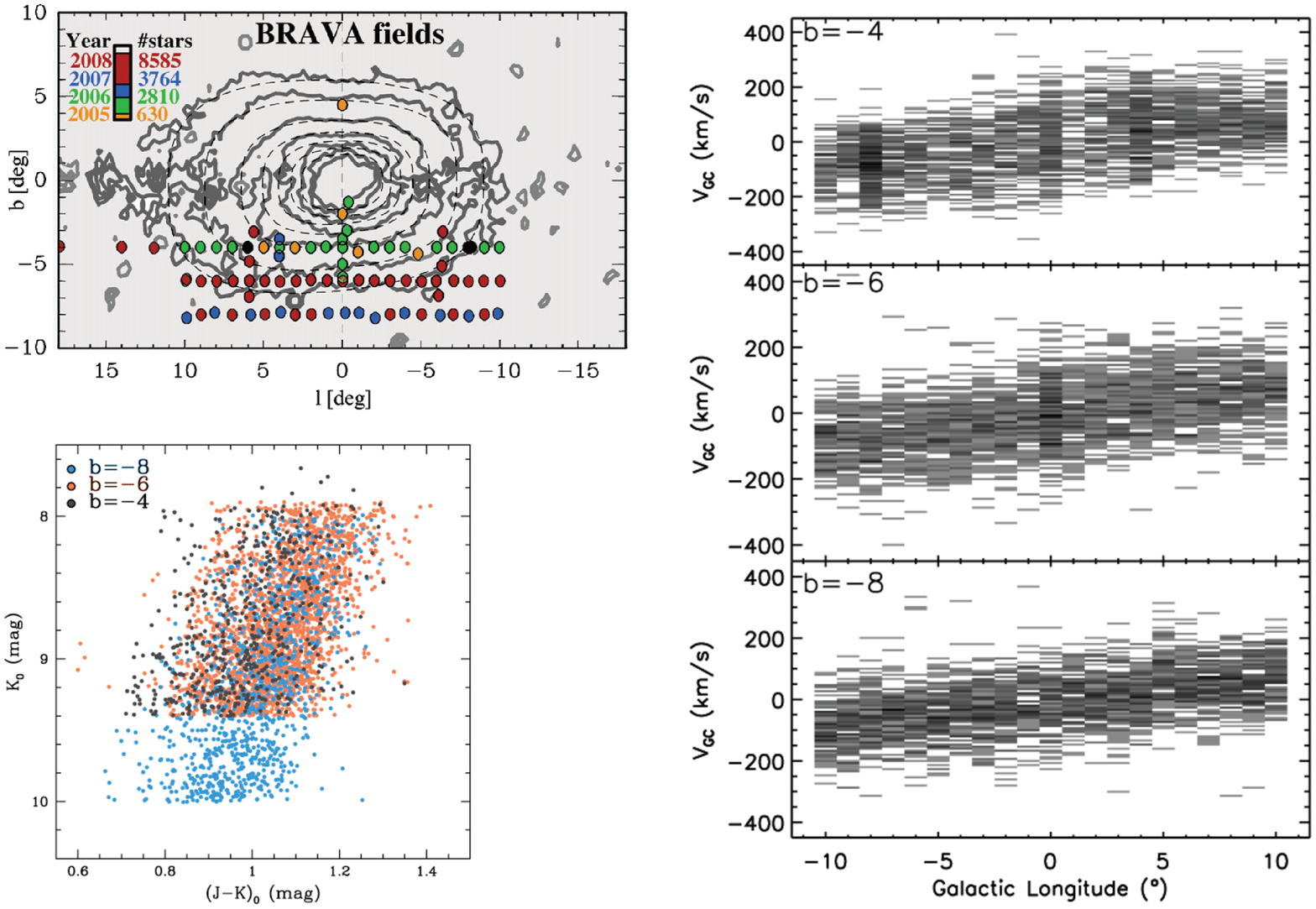}{bravaresults}{Summary of results from the Bulge Radial Velocity Assay (BRAVA); Rich et al.\ 2007; Kunder et al.\ 2012. Upper left:  BRAVA fields observed with Hydra, superimposed on the 4 $\mu \rm m$ isophotes of Launhardt et al.\ (2002).  Lower left: $K, J{-}K$ color-magnitude diagram of targeted red giants sorted by Galactic latitude. Right: The $l-v$ greyscale plot of the rotation curve; notice the decrease in dispersion with increasing Galactic latitude; also the rotation curve is invariant as a function of latitude; this is the ``cylindrical'' rotation indicative of the bar kinematics. }

\section{New Results from the BAaDE Survey:  Very Luminous AGB Stars with a High Velocity Dispersion}

The Bulge Asymmetries and Dynamics Evolution (BAaDE; principle investigator Y. Pihlstrom) survey employs a novel method developed by Lorant Sjouwerman (Sjouwerman et al.\ 2009; Sjouwerman et al.\ 2017) to discover and obtain radial velocities for SiO maser sources in the bulge and disk.  SiO masers are
mass-losing AGB stars; the presence of dust in the envelope creates an intense IR radiation field that facilitates the level inversions in the molecules, leading to the nonthermal masing effect.  Because detection is in the radio, extinction has no effect and the kinematics can be studied down to the Galactic plane; hence the distribution of these sources is unlike any other sample, yielding interesting kinematics inaccessible to other surveys (Figure \ref{baadefields}).   Following a selection from the MSX mid-IR color-color plane, the radial velocities can be rapidly surveyed with chains of maser sources being used for self-calibration.  The color selection has boosted the detection rate to well over 50\%, remarkably high for such studies.  This has vastly increased the numbers of masers, with roughly ${\sim}2700$ in Trapp et al.\ (2018); the numbers are likely to ultimately grow to over 20,000. 
We can understand the nature of the population by matching the radio emission with the 2MASS photometry. 

\articlefigure[width=0.95\textwidth]{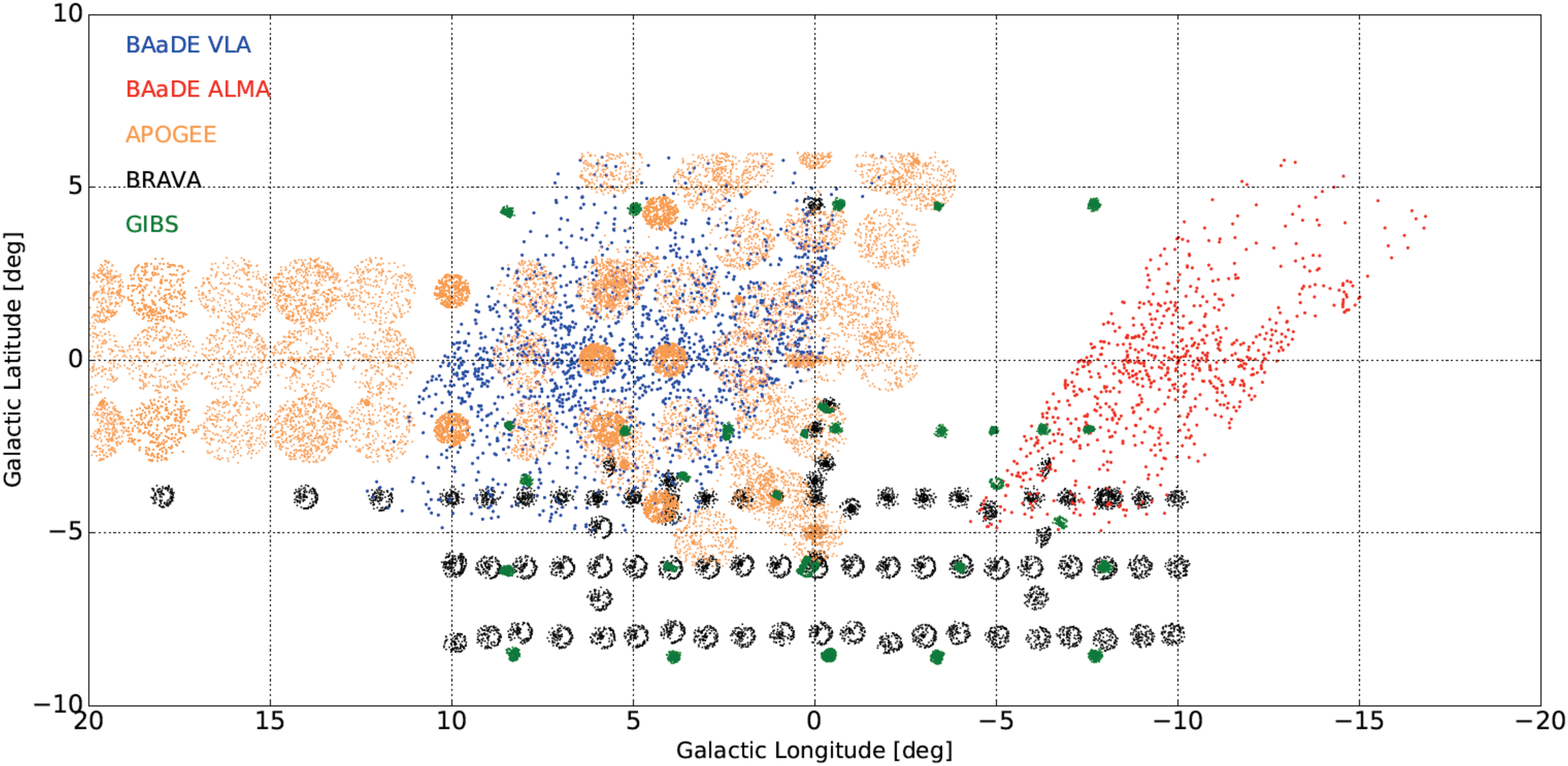}{baadefields}{Footprints of actual targets from optical/infrared bulge surveys with public data, and the BAaDE survey.  Notice that the BAaDE sources are found throughout the plane reaching $b=0^{\circ}$; detection in the radio is unaffected by extinction.  The ALMA sources are at $-5 < l <-25$.  The gap in the survey coverage at $0<l<-5$ has been observed with VLA; the data in this gap are presently under reduction; Stroh et al. 2019 in prep.}

The AGB population of the bulge is of {\it great interest} because it represents a present day snapshot of the progeny of the bulge turnoff stars.  It is also a potentially independent measure of whether some part of the bulge is intermediate age, but the results are complicated.   And recall that the main sequence turnoff is yielding a picture that is very far from definitive.   The SiO masers are primarily a luminous, red ($2 < J{-}K\,< 8$) population; there is overlap with the Mira variable population.   We can infer from the bolometric luminosities (for both Miras and the masers) whether the progenitors are old or intermediate age.  However this can be complicated for metal-rich stars (Guarnieri et al.\ 1998) which can reach high luminosity and still be old.  The Mira variables in Catchpole et al. also complicate matters: the long-period ${>}400$ day Miras tend the follow the bar, while the short-period Miras do not.  Does this suggest that at least part of the bar is intermediate age?

\section{Two Populations of SiO Masers}

We now turn to the SiO masers; this work is described in depth in Trapp et al.\ (2018).  
We began the project with only maser detections (for a given RA, DEC) and radial velocities; we had to solve the puzzle of how the SiO maser population fits into our understanding of the bulge.  This required matching of the SiO maser detections to 2MASS to produce a color-magnitude diagram. The breakthrough came when the velocity dispersion of the masers was tracked across the color-magnitude diagram (Figure \ref{apogeefig}).  The velocity dispersion doubles for stars with $K>5.5$, a clue that the brighter masers might lie in the foreground disk, and the fainter population in the bulge.    Trapp et al.\ (2018) suspected that this would be the case, but sought confirmation in the APOGEE experiment results.  One can sort the APOGEE sample by $\log g$ and see clearly a similar pattern:  the stars with lowest gravity (the bulge AGB stars) show the clear bulge kinematics pattern with a high velocity dispersion; the higher gravity stars have cold kinematics, presumably belonging to the disk.

As discussed in Trapp et al.\ (2018), four arguments are advanced to support the assertion of disk membership for the ``cold'' kinematic population, the last of which is a simple disk model that  reproduces the slope of the rotation curve.  

\articlefigure[width=1.0\textwidth]{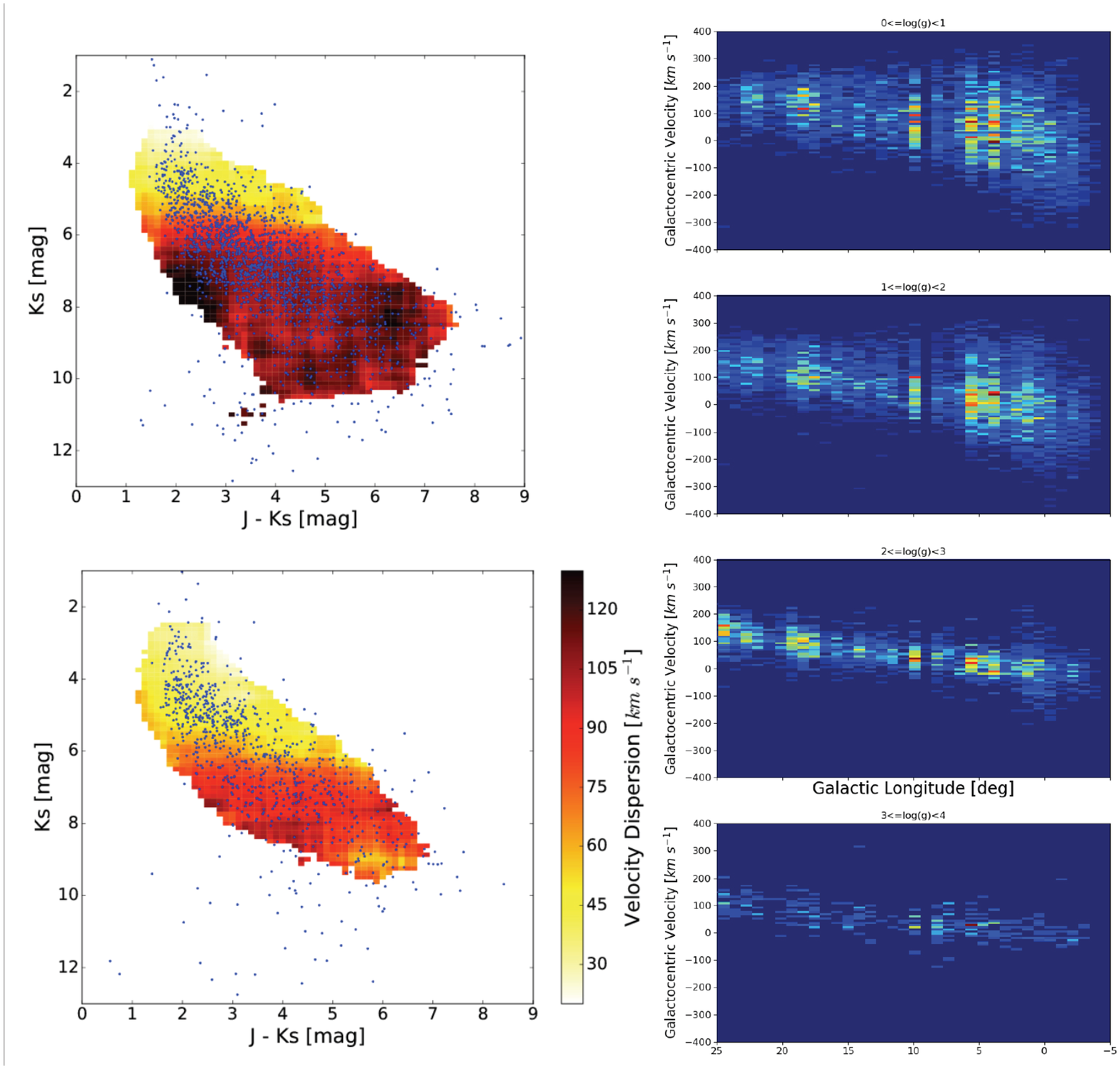}{apogeefig}{In the left pair of figures, we illustrate how the velocity dispersion varies across the $K,J-K$ color-magnitude diagram (with velocity dispersion heat scale; see Trapp et al.\ 2018 for details) for masers detected by VLA (top left) and ALMA (bottom left); notice the very sharp demarcation near $K=6$; stars fainter than this are in the bulge. The bar is more distant at negative latitudes.  The righthand panel shows the velocity dispersion of stars toward the bulge with $(J-K)>0.3$ and $K>5$, retrieved from APOGEE.   These stars are sorted by $\log g$ from low (giants; top) to high (subgiants; bottom).  The stars with lower velocity dispersion are tagged by APOGEE as being closer to the Sun; we propose that the low velocity dispersion SiO masers belong to the disk. }

Turning to the kinematics and spatial distributions in Figures \ref{siokinematics} and \ref{siogalactic}, the wisdom of applying the kinematic separation implied by the kinematic analysis of the color-magnitude diagram in Figure \ref{apogeefig} becomes evident.   A strikingly clear demarcation between the disk and bulge population is evident; indeed the SiO masers with the faintest $K$ magnitudes may be on the far side of the bar.  The concentration of the kinematic ``hot'' population masers to the plane is also clearly seen in Figure \ref{siogalactic}; it is reasonable to associate this population with the bulge.   

\articlefigure[width=1.0\textwidth]{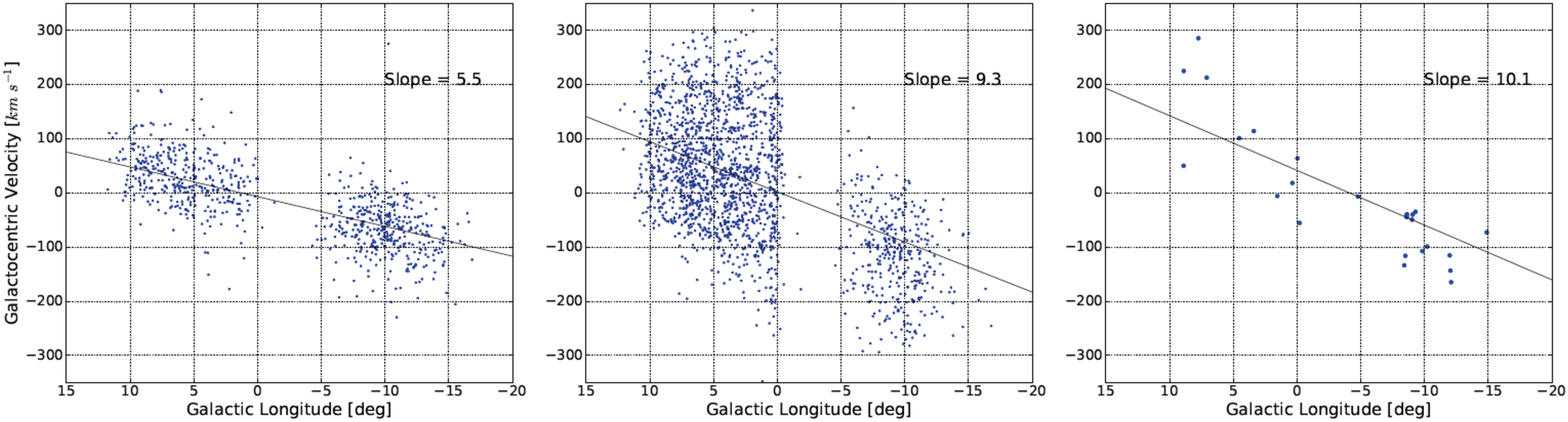}{siokinematics}{Kinematics of the SiO masers; the separation into three kinematics was achieved as described in the text and Figure 3.  Those masers with$K<6$ are the "cold"kinematic population (lefthand plot);  $K>6$ comprise the hot kinematic subpopulation (middle plot) with the faintest stars possibly lying on the far side of the bar (righthand plot).  See Trapp et al. 2015 for a full description. }

\articlefigure[width=1.0\textwidth]{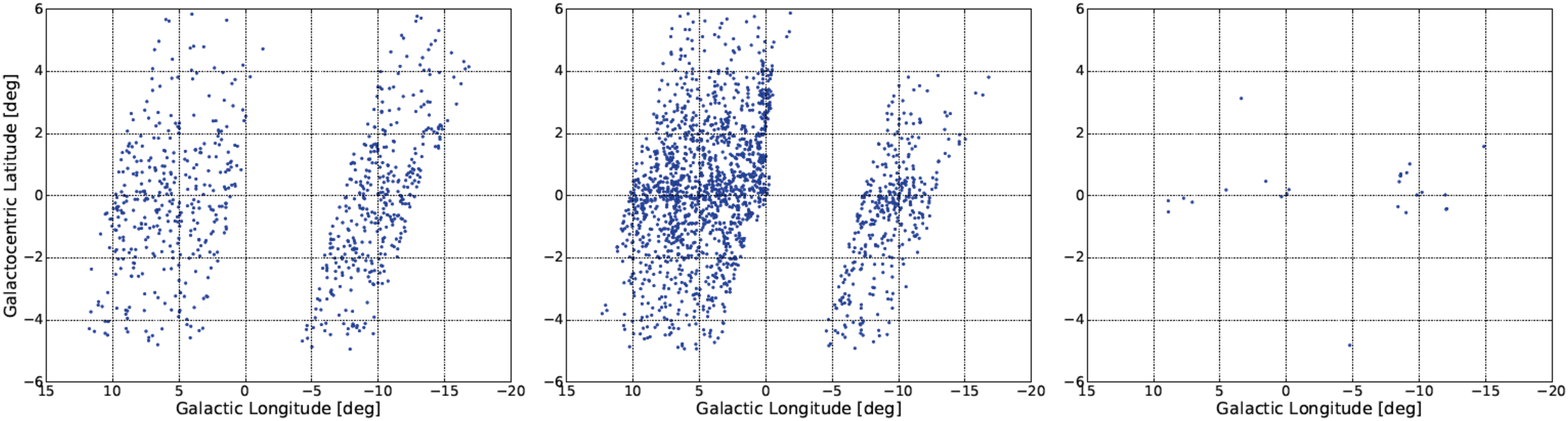}{siogalactic}{As in Fig \ref{siokinematics} but the sources are now shown in $l,b$.  Notice that the bulge sources show greater concentration to the plane, with the candidate far side sources having the tightest concentration to the plane.}

\subsection{Luminous AGB Stars in the Hot Kinematic Population}

Adopting a distance of 8.3 kpc to the Galactic Center (Gillessen et al.\ 2017)  and applying the bolometric correction of Messineo (2004), we can estimate the bolometric magnitudes in Figure \ref{mbolfig}.  We have obtained the distances to the putative disk population using a Monte Carlo population of a model disk.  This yields the unexpected result that the bulge population is bolometrically more luminous than that of the disk.  This is an unexpected and somewhat distressing result, suggesting that the bulge/bar has a greater fraction of intermediate age stars than the foreground disk.   The same figure shows that the most bolometrically luminous stars in the kinematically hot population are concentrated to the Galactic plane, while cold kinematic (disk) stars appear to show no such tendency.   The possibility of intermediate-age stars being concentrated toward the Galactic center is actually a well known result; the concentration of the most luminous AGB stars to the central 2$^\circ$ was established with the pioneering infrared scans of Catchpole, Whitelock, \& Glass (1990); Blommaert \& Groenewegen (2007) also found the longest period Mira variables concentrated toward the central 0.1$^\circ$.   It is therefore not too surprising that the SiO maser AGB population exhibits the same trend.    This finding calls for more investigation, especially a consideration of the kinematics and bolometric luminosity as a function of Galactic latitude; we are currently undertaking such an investigation by matching the SiO masers with Miras discovered via the Matsunaga's (2018) survey using a 1.5 m telescope in South Africa  (Rich et al.\ 2018, in preparation).   The extremely high luminosities are a concern; the stars are concentrated to the plane, but we are undertaking very dedicated efforts to re-evaluate the reddening, distance constraints from parallax, and the construction of new spectral energy distributions (SEDs) that will span from near IR to 20 $\mu \rm m$.   R. Sahai is presently involved in this effort.  Our team also plans employ new \emph{Gaia} derived distances to the disk population; Quiroga-Nu{\~n}ez et al. (2018).
It is difficult not to associate those stars at $M_{\rm bol}<-6$ with some kind of intermediate age to young stellar population; one cannot appeal to geometric depth to explain the high luminosities; incorporating a population of luminous giants in an actual dynamical model of the inner galaxy will be of significant value.   It will be important to bring to bear every resource in verifying these extreme luminosities.

\articlefigure[width=1.0\textwidth]{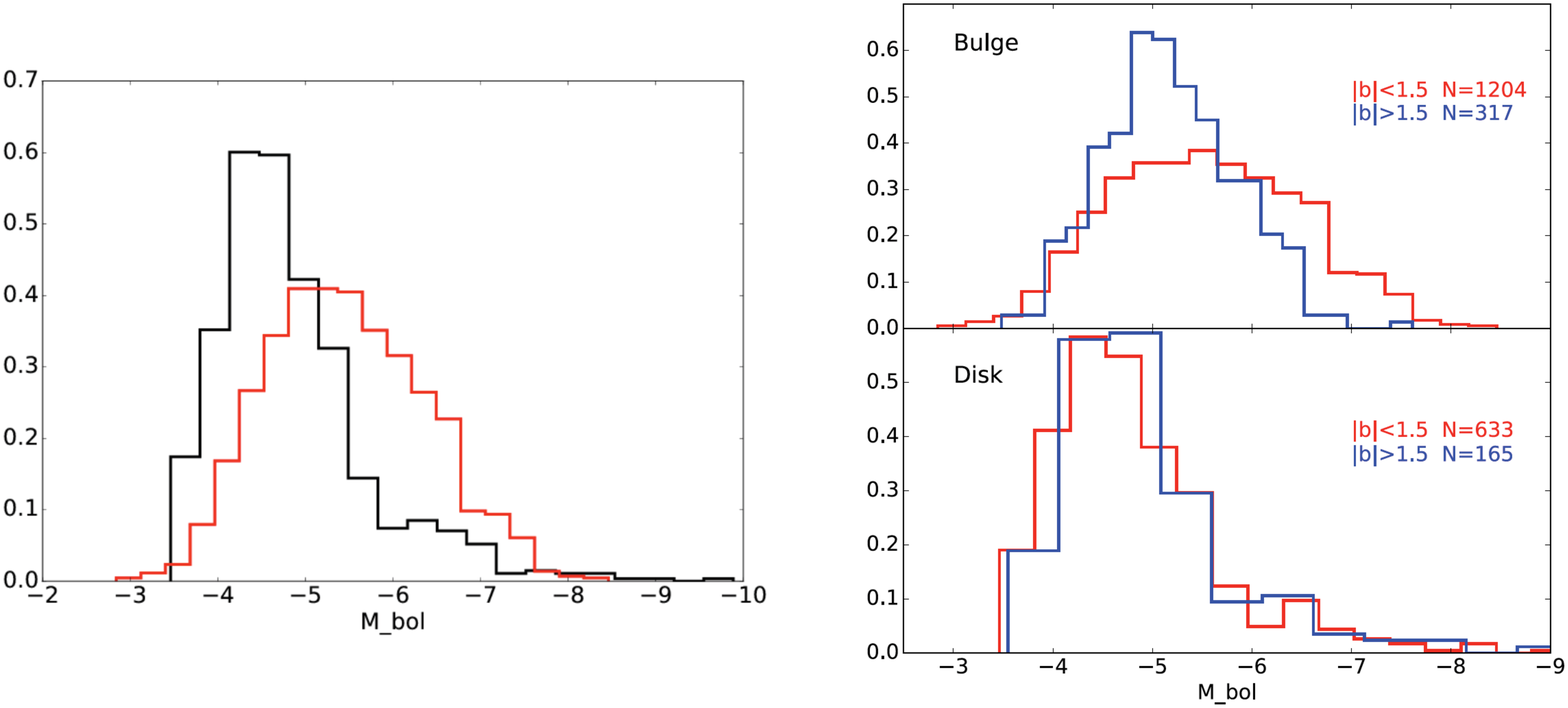}{mbolfig}{Distribution of bolometric magnitude by kinematic population.   In the {\it lefthand} panel, the red histogram is the kinematic hot (presumed bulge) population; the black histogram is the kinematic cold (presumed disk) population.  We have adopted a distance of 8.3 kpc for the ``bulge'' giants.  Notice the substantial number of bulge stars with $M_{bol}<-6$; the extremely high luminosities are a concern; our group is integrating the SEDs to test the high luminosities.   The disk stars are all assigned a distance of 3.4 kpc. {\it  Righthand }panel:  Here the red histogram indicates lower Galactic latitude.  When both the hot and cold kinematic populations are plotted as a function of Galactic latitude, the most luminous bulge (hot kinematic population) sources appear to be concentrated to the plane, while the cold kinematic population shows no tendency for the most luminous members to be confined to the disk.}

Figure \ref{dispersionfig} shows the almost perfect similarity between our maser population and the red clump stars of Babusiaux et al.\ (2014).  Red clump stars are core helium burning stars corresponding to the horizontal branch in metal-rich stellar populations.  Since that phase can only occur following the helium flash (implying red giants with degenerate cores as progenitors) we know that the red clump is at least 1 Gyr old, and likely older.  Indeed, the red clump is present throughout this region (see, e.g., Figer et al.\ 2004) and there is strong indication for a predominantly old stellar population in the Galactic Center (Pfuhl et al.\ 2011).    However, there is also evidence of significant star formation, especially in the disk.  It is difficult to reconcile the large radial velocity dispersion with the ${\approx} 1$ Gyr old AGB progeny of a hypothetical population formed in the disk.   

\articlefigure[width=1.0 \textwidth]{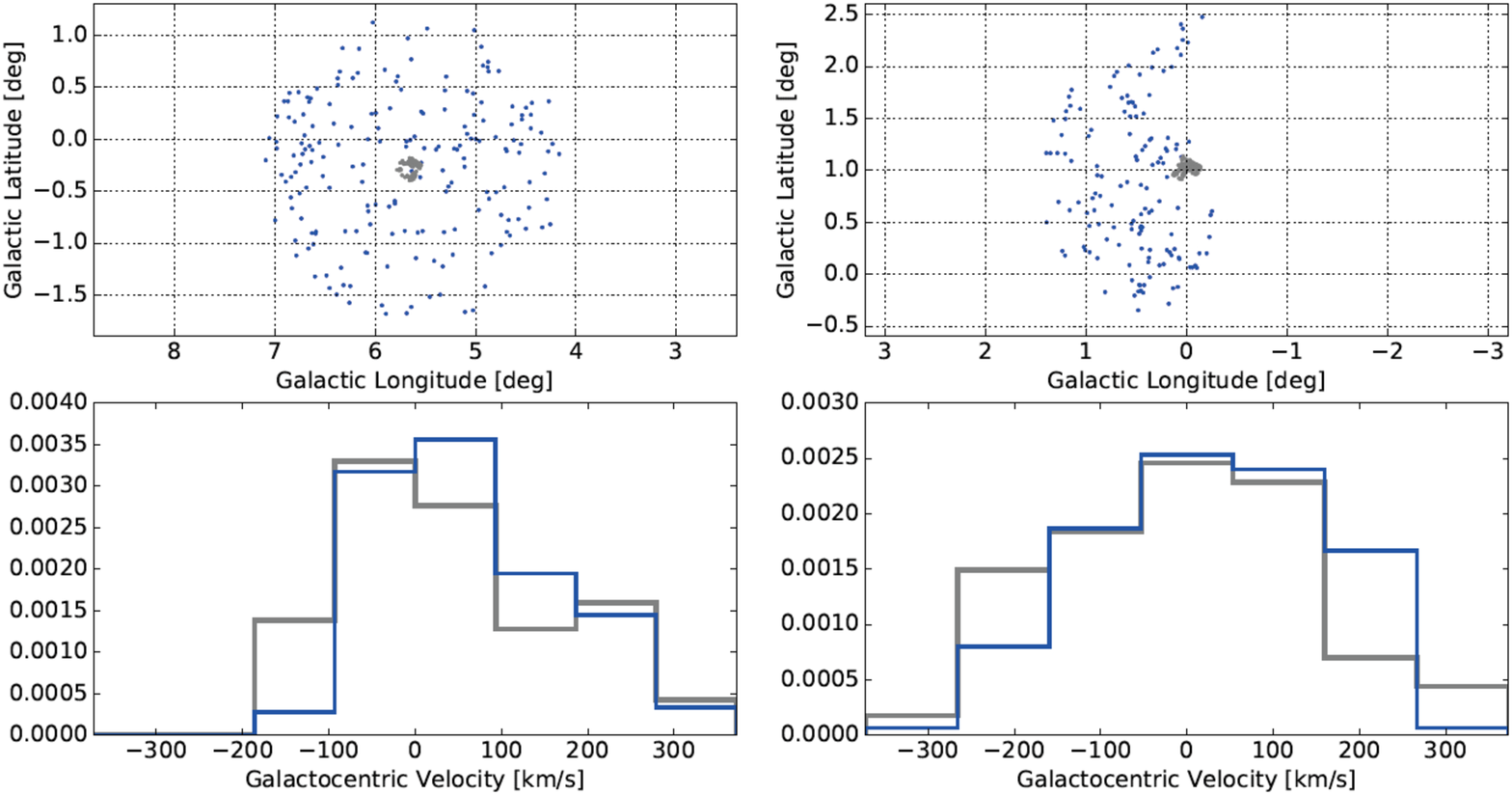}{dispersionfig}{We compare the velocity dispersion of a subset of our kinematic hot (presumed bulge) population maser sources with the red clump stars of Babusiaux et al.\ (2014).
The upper panels show the spatial distribution; the compact clump indicates the red clump sources.   Lower panel:  The distributions agree by the KS test, and the sources near $(l,b) =0$ have $\sigma = 145\pm 8 \rm km/sec$.   This is near the
maximum velocity dispersion seen the bulge (Valenti et al.\ 2018).  If the stars are only ${\sim}\,1-2$ Gyr old, how can they so rapidly acquire such a high velocity dispersion? }

\section{Conclusions}
We have traced a brief history of how the seminal Mould (1983) paper helped to inspire large scale kinematic studies of the Galactic bulge using late type giants as kinematic probes.  The BRAVA survey approach of characterizing kinematic  populations via the $l-v$ plot was used to attain an understanding of the BAaDE SiO maser population which has only radial velocities and matched infrared photometry.

The comparison of the BAaDE sample with other optical/infrared kinematic surveys, especially BRAVA, enables us to identify a candidate disk and bulge population of masers; this suspicion will be explored in greater precision using the
\emph{Gaia} database.  Assuming a common distance of 8.3 kpc, the hot kinematic population of masers we associate with the bulge reaches surprisingly high luminosities, up to $M_{bol}\approx -7$; further, the velocity dispersion of this population is identical to that of the presumably old red clump stars.   It would be difficult to attribute all of the luminous AGB stars to binary star evolution.  In a new result, we find that the hot kinematic population reaches high luminosities within $1.5^\circ$ of the plane, while the cold kinematic population shows no change in its luminosity distribution with respect to Galactic latitude.  This difference adds more strength to the argument that the most luminous AGB sources are intermediate age.  However, it remains surprising that these very luminous AGB stars show pure bulge kinematics, and the kinematics shows no indication (e.g. low velocity dispersion) oftheir  having recently been born in the disk.  The tension between evidence for a very old bulge and younger populations in the bulge continues.

The BAaDE collaboration is currently undertaking work that will improve our understanding of the population.  We are constructing multiwavelength SEDs for all of our sample, which will lead to more precise bolometric magnitudes.  We are also matching our sample to the \emph{Gaia} database; this has the potential to yield actual parallaxes to the disk sample, which is surprisingly bright and not reddened.   Even success with a subset will help, as will the proper motions for a much larger subset of our sample.  Following these efforts, it will be possible to undertake a better comparison with models and a deeper approach to the problem of what these luminous AGB stars reveal about the star formation history of the Galactic bulge.

\acknowledgements   
The BAaDE project is funded by National Science Foundation Grant AST-1517970.   R.M.R. also acknowledges
partial support from AST-1518271.

This paper uses data products obtained with instruments run by the National Radio Astronomy Observatory (NRAO): the VLA and ALMA. The NRAO is a facility of the National Science Foundation operated under cooperative agreement by Associated Universities, Inc. This research made use of data products from the Midcourse Space Experiment. Processing of the data was funded by the Ballistic Missile Defense Organization with additional funding from the NASA office of Space Science.
This research has also made use of the NASA/IPAC Infrared Science Archive, which is operated by the Jet Propulsion Laboratory, California Institute of Technology, under contract with the National Aeronautics and Space Administration.
 This publication makes use of data products from the Two Micron All Sky Survey, which is a joint project of the University of Massachusetts and the Infrared Processing and Analysis Center/California Institute of Technology, funded by the National Aeronautics and Space Administration and the National Science Foundation.



\end{document}